\documentclass[reprint,amsmath,amssymb,aps,superscriptaddress,nofootinbib]{revtex4-2}
\usepackage{graphicx}
\usepackage{dcolumn}
\usepackage{bm}
\usepackage[utf8]{inputenc}
\usepackage{amsmath, amsthm, amsfonts, amssymb}
\usepackage[svgnames]{xcolor}
\colorlet{color1}{NavyBlue}
\usepackage[colorlinks=true,allcolors=color1]{hyperref}
\usepackage{physics}
\usepackage{dsfont}
\usepackage{graphicx}
\usepackage{cleveref} 
\usepackage{xfrac}
\usepackage{mathrsfs}

\begin{document}
	
	\title{Classical mass inflation vs semiclassical inner horizon inflation}
	
	\author{Carlos Barceló}
	\email{carlos@iaa.es}
	\affiliation{Instituto de Astrofísica de Andalucía (IAA-CSIC), Glorieta de la Astronomía, 18008 Granada, Spain}
	\author{Valentin Boyanov}
	\email{vboyanov@ucm.es}
	\affiliation{Departamento de Física Teórica and IPARCOS, Universidad Complutense de Madrid, 28040 Madrid, Spain}
	\author{Raúl Carballo-Rubio}
	\email{raul@sdu.dk}
	\affiliation{CP3-Origins, University of Southern Denmark, Campusvej 55, DK-5230 Odense M, Denmark}
	\affiliation{Florida Space Institute, 12354 Research Parkway, Partnership 1, Orlando, FL 32826-0650, USA}
	\author{Luis J. Garay}
	\email{luisj.garay@ucm.es}
	\affiliation{Departamento de Física Teórica and IPARCOS, Universidad Complutense de Madrid, 28040 Madrid, Spain}
	\affiliation{Instituto de Estructura de la Materia (IEM-CSIC), Serrano 121, 28006 Madrid, Spain}
	
	\begin{abstract}
		We analyse the geometry of a spherically symmetric black hole with an inner and outer apparent horizon which is perturbed by spherical null shells of matter. On a classical level we observe that the mass inflation instability is triggered, resulting in a growth of curvature and an inward displacement of the inner horizon. We study in detail the inner structure of the mass-inflated region and compare it with previous results obtained for the case in which the perturbing matter content has a continuous distribution. We then perform an approximate calculation of the renormalised stress-energy tensor of a quantum field in the vicinity of the inner horizon, and analyse the semiclassical backreaction on this region of the geometry. We find that the classical tendency for this horizon to move inward due to mass inflation is challenged and potentially overcome by a semiclassical tendency for it to inflate outward.
	\end{abstract}
	
	\maketitle
	\section{Introduction}

	A generic outcome of gravitational collapse is the formation of a trapped region, and a generic characteristic of trapped regions is the presence of both an outer and an inner apparent horizon. In classical General Relativity, strict spherical symmetry, and no charge, any inner apparent horizon would rapidly disappear into a singularity. However, the presence of the slightest amount of angular momentum or electric charge, an effective regularized central region, or even just a seemingly innocuous transient phenomenon, can serve to breathe additional life into this horizon~\cite{Wald1984,Hayward,Ansoldi}.
	
	A long-lived inner horizon leads to the well-known mass inflation instability, whereby small perturbations in the matter content of the geometry result in a highly non-linear response in the increase of curvature~\cite{PoissonIsrael89}. For the charged black hole and many other models, this increase in curvature can be related to a growth of the Misner-Sharp mass \cite{Misner1964}, hence the name ``mass inflation". For an initially static geometry with an inner horizon which is then perturbed, the region where mass is ``inflated" begins close to the initial position of this horizon and extends below it. Marking the beginning of this large-curvature region is a shockwave \cite{Marolf2012} located on a null hypersurface which remains in the vicinity of the initial inner horizon position. Meanwhile, the inner horizon moves inside this region along a timelike hypersurface, tending toward the origin. If it reaches the origin before reaching the Cauchy horizon, a spacelike singularity is formed, as observed in the numerical analysis in \cite{Brady1995}, in addition to the null weak singularity at the Cauchy horizon itself \cite{Ori1991,Ori1992,Ori1998,Dafermos2017}.\par
	
	Classically, this entire process is hidden behind an event horizon and has no effect on the outside universe. Only observers who enter the black hole may worry about whether they will survive their altogether turbulent journey and what, if anything, may await them beyond the Cauchy horizon. However, the picture is quite different when quantum effects \cite{BD,Wald1995} start to be taken into account. To begin with, the energy content of the vacuum state of quantum fields which reside on these spacetimes tends to a divergence at the Cauchy horizon, and does so with enough strength to rule out the possibility of extensions of spacetime beyond this horizon in a semiclassical regime \cite{Birrell1978,BalbinotPoisson93,Ori2019,Hollands2020a,Hollands2020b,Zilberman2022}. Additionally, for a dynamically formed black hole, if one takes into account Hawking evaporation of the outer horizon~\cite{Hawking1975,Hawking1974}, then the trapped region should disappear long before a Cauchy horizon or a null singularity forms (assuming no stable extremal remnant is generated).\par
	
	Furthermore, as shown in a previous work by the present authors \cite{Barcelo2021}, the backreaction from the quantum vacuum around the inner apparent horizon (i.e. the inner boundary of a dynamically formed trapped region) should also be taken into account long before any consideration of the Cauchy horizon and its divergences. Particularly, we used the renormalised vacuum expectation value of the stress-energy tensor operator (which we will refer to as just the ``renormalised stress-energy tensor", or RSET), constructed from a quantum massless scalar field and calculated in the Polyakov approximation for spherical symmetry, to analyse semiclassical backreaction perturbatively around the inner horizon. What we found is a general tendency for the perturbed inner horizon to move outward, reducing the size of the trapped region. This movement is analogous to the Hawking evaporation from the other side of the trapped region, as both are sourced by negative energy fluxes with magnitudes related to the surface gravities of their respective horizons. The main difference between the two is the fact that a quasi-stationary approximation cannot be used for the inner horizon, as backreaction tends to have a quicker accumulated effect there. In the same sense as the term \textit{evaporation of the outer horizon} conveys an idea of slowness, an adequate term for its interior counterpart might be \textit{inflation of the inner horizon}, recalling its more explosive tendency.\par
	
	It is the goal of this work to derive a similar perturbative semiclassical backreaction scheme on top of classical backgrounds which capture more realistically the evolution of the inner apparent horizon, i.e. backgrounds which represent classical mass inflation. To this end, we construct a simple geometric model with a spherically symmetric black hole perturbed by an outgoing null shell travelling between its outer and inner horizons (akin to Ori's model \cite{Ori1991}), and by a series of ingoing null shells with decreasing mass, which represent the power-law decay of perturbations typically considered for mass inflation \cite{PoissonIsrael89}. Then, by deciding what mass and charges each shell carries and how these affect the position of the inner horizon, we analyse the possibilities for the resulting evolution of the trapped region and the global causal structure of the spacetime.\par
	
	On top of these spacetimes we construct a quantum ``in" vacuum state and calculate the RSET in the Polyakov approximation in the form of a series expansion around the inner horizon. We once again find a tendency for the semiclassical source to move the inner horizon outward, although in this case it competes with the classical tendency for it to move inward due to mass inflation. As the inner horizon moves inward, the classical term which drives it to do so decreases, while the semiclassical term which pushes it outward increases, eventually overcoming its Planck-scale suppression and dominating over its classical counterpart when the surface gravity of this horizon becomes Planckian.\par
	
	This analysis of initial tendencies may seem to suggest that semiclassical effects can only dominate when the spacetime curvature becomes Planckian (which in itself is an important result), but a full semiclassically self-consistent solution obtained from a family of particularly simple mass inflation geometries reveals than this need not be the case. Terms sourced by the RSET which drive the slow-down and reversal of the inner horizon tendency toward the origin can in fact grow exponentially quickly, and lead to this reversal while curvature is still far from being Planckian. This solidifies the conclusion that horizon-related semiclassical effects play an important role for the evolution of a generic black-hole trapped region.\par
	This work is structured as follows. In section \ref{secshells} we construct our shell-based classical perturbation model which triggers mass inflation on an initially static spherically symmetric black hole. In section \ref{secri} we analyse in detail the inner regions of the resulting geometries, focusing in particular on the evolution of the inner apparent horizon. We also show the global causal structure which the purely classical evolution leads to, with its corresponding weak and strong singularities. Then, in section \ref{secsemicl}, we calculate the RSET in the Polyakov approximation for these spacetimes and tackle the backreaction problem. In section \ref{secconcl} we conclude with an overview of our results and we discuss the changes they may imply for the standard picture of black hole formation and evaporation.
	
	\section{Mass inflation with thin shells}\label{secshells}
	
	The model we will work with is a spherically symmetric geometry with a line element
	\begin{equation}
		ds^2=g_{tt}dt^2+g_{rr}dr^2+r^2d\Omega^2,
	\end{equation}
	where $d\Omega^2$ is the line element of the unit sphere. The simplicity of our construction lies in considering that this geometry is static, i.e. that the metric components $g_{tt}$ and $g_{rr}$ are functions of $r$ only, in patches separated by spherical null shells.
	
\begin{figure}
		\centering
		\includegraphics[scale=.8]{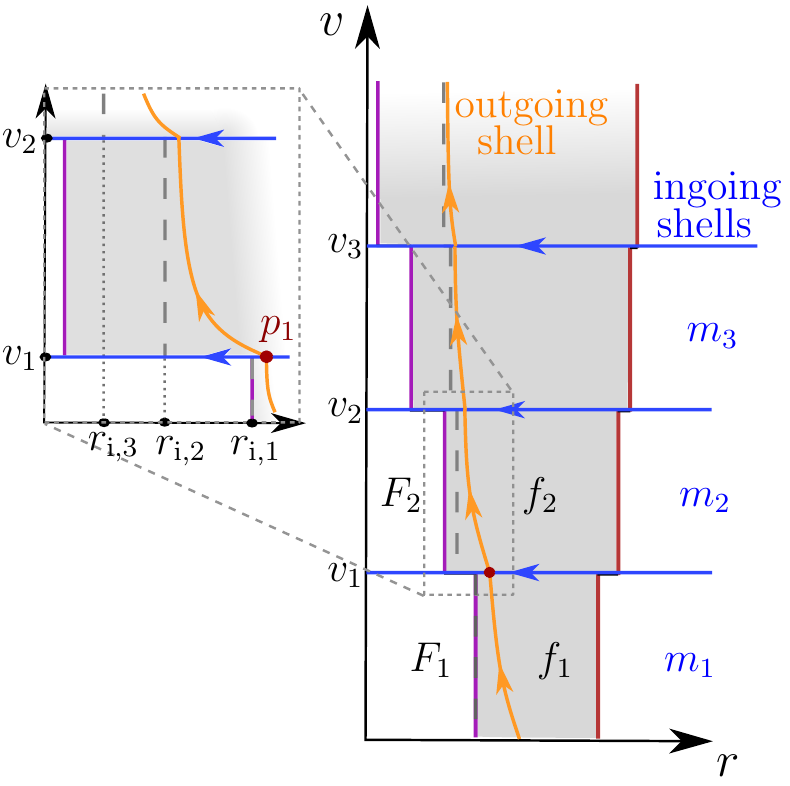}
		\caption{Static black hole with outer and inner horizons perturbed by an outgoing shell and a series of ingoing shells with decreasing mass, represented in advanced Eddington-Finkelstein coordinates. The trapped region is shaded in grey. The ingoing shells are located at $v_n$, with $n=1,2,\dots$; $m_n$ refers to the Bondi mass in the asymptotic region before the corresponding ingoing shells (outside the outgoing one); $f_n$ is the $g^{rr}$ metric component in each of these regions down to the outgoing shell, while $F_n$ is the same metric component on the inside of the outgoing shell. $r_{{\rm i},n}$ are the zeros of the $f_n$ functions extended to the region below the outgoing shell (they are the radii which this shell approaches exponentially in $v$ in each patch).}
		\label{f1}
	\end{figure}

	To begin with, consider that there are two null shells, one outgoing and one ingoing, which intersect at a point $p_1=(t_1,r_1)$. Continuity of the metric imposes the relation (see \cite{thooft1985,Redmount1985,Barrabes1990})
	\begin{equation}\label{intersect}
		\left|\frac{f_{2}}{f_{1}}\right|=\left|\frac{F_{2}}{F_{1}}\right|\quad \text{at}\quad p_1,
	\end{equation}
	where the functions $F$ and $f$ are the $g^{rr}$ component of the metric in different patches, with upper (lower) case letters referring to the interior (exterior) of the outgoing shell and the subscript ``$1$" (``$2$") referring to the interior (exterior) of the ingoing shell, as shown schematically in fig.~\ref{f1}.
	
	Now consider that this geometry has a trapped region (i.e. a region where $g_{tt}$ becomes positive and $g_{rr}$ negative, which directly relates to the expansion of outgoing null geodesics congruences becoming negative \cite{Hayward1994}). For simplicity, consider also that $-g_{tt}=g^{rr}$ in the static coordinates of each patch in the vicinity of the shells (as would be the case for e.g. the charged black hole, and many regular black hole models \cite{Hayward,Ansoldi}). We will refer to this metric component as the \textit{redshift function}, this being the previously defined $f$ (and $F$).\footnote{Throughout the text, it should be recalled that the properties we require from this function are related to both $g_{tt}$ (which often bears the name redshift function by itself) and $g^{rr}$; the former has more to do with the trajectories of null geodesics, while the latter has to do with the junction conditions which drive the dynamics of the spacetime.} We take the outgoing shell to be travelling inside the trapped region, between the outer and inner horizons. Once we specify some aspects of the initial conditions in the patches of $F_1$, $f_1$, and $f_2$, which amounts to choosing the mass and charge carried by each shell, eq. \eqref{intersect} will tell us how the redshift function in the future of the innermost region of the black hole, $F_2$, behaves. We will then repeat the process by adding more ingoing shells, as shown in fig. \ref{f1}, representing an ingoing perturbation of decreasing amplitude, and see when and how mass inflation is triggered.
	
	In advanced Eddington-Finkelstein coordinates we can write the line element in the $f_1$ patch as
	\begin{equation}\label{geo1}
		ds_{1}^2=-f_{1}(r)dv^2+2dvdr+r^2d\Omega^2.
	\end{equation}
	We note that the same coordinate $v$ can be used for all static representations of the patches external to the outgoing shell (while for those on the inside of this shell we will use a coordinate $V$). This outgoing shell is travelling along a null geodesic for both the geometries in the $f_1$ and $F_1$ patches. For now, we will describe its dynamics and its interactions with the ingoing shell(s) in the coordinates of the external $f_1$ patch, while in the next section we will go into more detail regarding its description from the point of view of the interior patches. Being inside the trapped region of \eqref{geo1}, its movement is described by the equation
	\begin{equation}
		\frac{dr_{\rm shell}}{dv}=\frac{1}{2}f_1.
	\end{equation}
	Since $f_1$ is negative inside the trapped region, the solution for its radial position $r_{\rm shell}$ decreases in $v$, eventually tending toward the inner gravitational radius of the $f_1$ patch, $r_{\rm i,1}$, where the redshift function can be approximated by
	\begin{equation}\label{f1exp}
		f_1(r)=-2\kappa_{\rm i}(r-r_{\rm i,1})+\mathcal{O}[(r-r_{\rm i,1})^2],
	\end{equation}
	with $\kappa_{1}$ being the absolute value of the surface gravity of this inner radius (i.e. $-\frac{1}{2}\partial_rf_1|_{r=r_{\rm i,1}}$). To clarify, $r_{\rm i,1}$ is the position the inner horizon would have if the $f_1$ patch were continued to below the outgoing shell (i.e. the zero of the $f_1(r)$ function when extended to below $r_{\rm shell}$). We note that since below $r_{\rm shell}$ the geometry is described by the $F_1$ function, the position of the actual inner apparent horizon of the geometry is in fact the zero of $F_1$. This position will be denoted by $R_{\rm i,1}$ later on.
	
	The solution which describes the tendency of the shell toward $r_{\rm i,1}$ at sufficiently large values of $v$ can be obtained from the linear order in the above expansion,
	\begin{equation}\label{rsell1}
		r_{\rm shell}(v)=r_{\rm i,1}+(r_0-r_{\rm i,1})e^{-\kappa_1 v}+\order{e^{-2\kappa_1 v}},
	\end{equation}
	where $r_0$ is a positive constant representing the position of the shell at $v=0$. Substituting \eqref{rsell1} into \eqref{f1exp}, the redshift function on the shell is then approximated by
	\begin{equation}\label{eq:f1_app}
		f_{1}|_{\rm shell}=-2\kappa_{1}(r_0-r_{\rm i,1})e^{-\kappa_{1}v}+\order{e^{-2\kappa_1 v}}.
	\end{equation}
	
	Now consider that this outgoing shell is intersected by the ingoing one. If the latter is assumed to have positive mass (more generally, satisfies the null energy condition), the inner horizon can only be displaced inward, i.e. $r_{\rm i,1}\ge r_{\rm i,2}$, following a timelike trajectory, in the same way as the outer horizon can only be displaced in an outward (spacelike) direction.\footnote{This is simply an extension of Hayward's theorem for continuous matter \cite{Hayward1994} to the case of shells. Intuitively, it can be seen from the fact that the inner horizon acts as a surface of accumulation of geodesics, which restricts its movement to causal directions.} Let us also assume that this mass which the shell carries to the black hole is small (compared to the total black hole mass), as one may expect from perturbation tail analyses in an astrophysical scenario. The redshift function after this shell can then be approximated by
	\begin{equation}\label{eq:f2_app}
		f_{2}|_{\rm shell}=-2\kappa_{2}(r_{\rm shell}-r_{\rm i,2})+\mathcal{O}[(r_{\rm shell}-r_{\rm i,2})^2].
	\end{equation}
	Particularly, the smallness of the mass carried by the shell is meant to ensure that the order $(r_{\rm shell}-r_{\rm i,2})^2$ remains negligible (in units of the characteristic length scale of the black hole, i.e. the initial inner horizon radius), in the same way as the higher order terms in \eqref{f1exp} are, which amounts to a requirement that the that the displacement $(r_{\rm i,1}-r_{\rm i,2})$ be small (which we will impose explicitly later on). At the intersection point $p_1$, we can insert \eqref{rsell1} for $r_{\rm shell}$ into eq.~\eqref{eq:f2_app} which, together with eq.~\eqref{eq:f1_app}, gives us the expression for one of the two quotients from eq. \eqref{intersect}:
	\begin{equation}\label{quotient}
		\left.\frac{f_2}{f_1}\right|_{r=r_1}=\frac{\kappa_2}{\kappa_1}+\frac{\kappa_2}{\kappa_1}\frac{r_{\rm i,1}-r_{\rm i,2}}{r_0-r_{\rm i,1}}e^{\kappa_1v_1}\left[1+\order{e^{-\kappa_1v_1}}\right],
	\end{equation}
	where $v=v_1$ corresponds to the position of the ingoing shell, and is thus also the intersection time. The exponential in this expression, along with eq. \eqref{intersect}, already indicates that the value of redshift function $F_2$ can become much greater than $F_1$, in a manner suggestive of mass inflation. Particularly, one can easily see how this growth of $F_2$ relates to an increase of mass in e.g. the Reissner-Nordström geometry (which we will use as an illustrative example throughout this work), where $F_n=1-2M_n/r+Q_n^2/r^2$. If the redshift function increases in absolute value, and since it is negative ($r_1$ begin inside the trapped region), it translates into an increase in the only negative term it contains: the mass of the black hole. Assuming this term is already large from previous shell crossings and dominates the behaviour of the redshift function $F_1$ close to the shell, we have the relation
	\begin{equation}
		F_{1}|_{r=r_1}=-\frac{2M_{1}}{r_1}+\order{M_1^0},
	\end{equation}
	where $M_1$ is the mass on the inside of the outgoing shell before $v_1$. With the same assumption for the region after the ingoing shell, i.e. that the mass term dominates in the redshift function, we get the inflated mass of the charged black hole
	\begin{equation}\label{reissner}
		M_{2}=M_{1}\left.\frac{f_{2}}{f_{1}}\right|_{r=r_1}+\order{M_1^0}.
	\end{equation}
	However, as the reader may have already noticed, the exponential in \eqref{quotient} has a prefactor which must also be carefully analysed. For example, if the ingoing shell is considered to have a particular charge and mass which make the displacement of the inner gravitational radius $(r_{\rm i,1}-r_{\rm i,2})$ sufficiently small, the exponential growth could be cancelled. We will discuss this in more detail in the following.\par
	Let us make the perturbation by ingoing shells an iterative process: we represent an ingoing, polynomially decreasing flux of radiation (which is usually the source of mass inflation \cite{PoissonIsrael89,Ori1991} stemming from the decay of perturbations 
	on the geometry \cite{Dafermos2016}) with a sequence of ingoing shells of progressively smaller mass,
	\begin{equation}\label{deltam}
		\delta m_{n}=\frac{a}{v_n^q},
	\end{equation}
	where $\delta m_{n}$ refers to the change of the exterior mass (i.e. the Bondi mass \cite{Bondi1962} related to past null infinity) produced by a particular ingoing shell located at $v=v_n$, the power $q$ is positive,\footnote{$q$ has a lower bound which depends on the spacing of the shells, needed to guarantee that the total mass thrown into the black hole is finite. For example, for a linear distribution of shells in $v$, $q>1$. If the shells become more spread-out, then $q$ can be smaller, while if they become more concentrated it must be larger.} and $a$ is a positive constant with appropriate dimensions. We also impose that there be infinitely many shells and that $\lim_{n\to\infty}v_n=\infty$.\par
	
	One of the key ingredients necessary for mass inflation is that this increase in the mass (as seen from outside the object) is itself related polynomially to the change in position of the inner gravitational radius $\delta r_{{\rm i},n}$, i.e.
	\begin{equation}\label{ingredient1}
		\delta r_{{\rm i},n}=-\frac{\beta}{v_n^p},
	\end{equation}
	where the power $p$ is again positive (though it can be different from $q$), and $\beta$ is again a positive constant (at least asymptotically in $v$). We stress that eq. \eqref{ingredient1} is an \textit{assumption} about how the geometry responds to the ingoing perturbation \eqref{deltam}. For example, in the Reissner-Nordström case, where one has
	\begin{equation}\label{rirn}
		\delta r_{{\rm i},n}=\frac{1}{\sqrt{m^2-Q^2}}(-r_{{\rm i},n}\delta m_n+Q\delta Q_n),
	\end{equation}
	the assumption \eqref{ingredient1} restricts the amount of (same sign) charge the ingoing shells can carry. One can easily imagine a case in which the relation between $\delta m_n$ and $\delta Q_n$ is such that e.g. $\delta r_{{\rm i},n}=0$, which would lead to an absence of mass inflation. This is not limited to the Reissner-Nordström case: \eqref{ingredient1} generally avoids the suppression of the exponential in \eqref{quotient} by its prefactor, allowing mass inflation to take place.
	
	This can be seen by looking at the evolution of the redshift function in the interior of the outgoing shell after the $n$th ingoing shell has crossed it, $F_n$. From eqs. \eqref{intersect}, \eqref{quotient}, and \eqref{ingredient1} we get the asymptotic relation at the $n$th intersection point
	\begin{equation}\label{massiteration}
		\begin{split}
			F_n&=F_{n-1}\frac{f_n}{f_{n-1}}\\&=F_{n-1}\frac{\kappa_n}{\kappa_{n-1}}\left(\frac{v_{n-1}}{v_n}\right)^pe^{\kappa_{n-1}\Delta v_n}\left[1+\order{e^{-\kappa_{n-1}\Delta v_n}}\right]\\&=F_{n-1}e^{\kappa\Delta v_n}\left[1+\order{e^{-\kappa\Delta v_n},\frac{\Delta v_n}{v_{n-1}}}\right],
		\end{split}
	\end{equation}
	where $\Delta v_n=v_n-v_{n-1}$ and in the last line we have directly substituted the asymptotic value of the surface gravity of the $f_n$ geometries, $\kappa_n\to\kappa$. We have also used the fact that the differences $v_n-v_{n-1}$ become negligible when compared to the absolute value of $v_n$ asymptotically (for any distribution of infinite shells which covers an infinite range in $v$), and that $e^{\kappa\Delta v_n}>1$, leaving only the dominant contribution as leading order. If we use this relation iteratively from an initial time $v=0$ when the interior redshift function was $F_0<0$, we get the asymptotically exponential increase (of the absolute value of) the redshift function at the shell
	\begin{equation}\label{massinfl}
		F_n|_{\rm shell}\sim F_0e^{\kappa v_n}.
	\end{equation}
	The result is independent of the spacing between the ingoing shells: increased spacing only leads to a higher jump in $F_n$ when a shell eventually falls. This tendency continues for as long as more shells are thrown in. Returning to the Reissner-Nordström case, we can once again easily associate this with a proportional increase of the mass term through eq. \eqref{reissner}. In more general geometries, it can be related to an increase of the Misner-Sharp mass given by \cite{Misner1964,Hayward1994}
	\begin{equation}\label{Misner-Sharp}
		M_{\rm MS}=\frac{1}{2}r(1-F),
	\end{equation}
	for this interior region. Let us recall that the Misner-Sharp mass provides a quasi-local characterization of gravitational energy in spherical symmetry~\cite{Hayward1994}.
	
	Eq. \eqref{massinfl} captures the main result of mass inflation: the exponential growth of the redshift function (more generally, the $g^{rr}$ metric component) below a certain radius, associated with a corresponding growth of the Misner-Sharp mass. In our model it can be physically interpreted in terms of the exchange of mass between the outgoing and ingoing shells. The outgoing shell, being inside the trapped region, can be seen as having a negative asymptotic mass. At the intersection points, the ingoing shells can therefore take away positive mass by making the negative one of the outgoing shell increasingly more so. This exchange is mediated by the dynamics of the gravitational field, and in particular its exponential nature is only triggered if the outgoing shell is taken from an initial proximity to the inner gravitational radius of the $f_n$ geometry patches, and subsequently (after the interaction) ends up deeper inside the trapped region due to the inward displacement of this inner gravitational radius \eqref{ingredient1}.
	
	It is interesting to note that this process is independent of the particularities of the infalling matter shells, only requiring that the perturbations induce the polynomially decreasing response of the inner gravitational radius \eqref{ingredient1}. If the shift in its position were instead to decrease, e.g., exponentially,
	\begin{equation}
		\delta r_{{\rm i},n}=-\tilde{\beta}e^{-\sigma v},
	\end{equation}
	where $\tilde{\beta}$ and $\sigma$ are positive constants, then one can observe from \eqref{quotient} and the corresponding equivalent of \eqref{massiteration} that the outcome would depend on the difference between the surface gravity $\kappa$ and the coefficient $\sigma$. Particularly, mass inflation would only take place if $\kappa>\sigma$, as has been shown in the case where $\sigma$ is the surface gravity of a cosmological horizon in asymptotically de Sitter spacetimes, governing the decaying tail of infalling radiation \cite{Mellor1992,Brady1992}.
	
	It is worth noting that this shell-based model has some characteristics in common with the case in which a continuous power-law flux of radiation is present, but there are also some differences. For example, the exponential which appears in the mass inflation of this model is directly related to the exponentially decreasing separation between the outgoing shell and the inner gravitational radius $r_{\rm i}$ (as seen from the outside) in each step represented in fig. \ref{f1}. Although the average of this distance taken over several steps has the same inverse-polynomial decrease as is expected from a continuous matter case (that is, if $r_{\rm i}$ had a position evolving as a continuous version of \eqref{ingredient1}), taking the limit to a continuous ingoing distribution of matter is not at all straightforward. Although the outgoing shell seems to be a good model for the shockwave which generally appears in this region even with continuous matter \cite{Marolf2012}, the ingoing shells offer qualitatively new features. This may also be inferred by the fact that a generalisation of Ori's model~\cite{Raul2021}, in which the ingoing flux is continuous, leads to a larger variety of asymptotic outcomes, unlike the single exponential behaviour observed here. The question of whether the continuous or the discrete model (or some combination of the two) fits best the behaviour of small (possibly quantised) infalling perturbations may thus turn out to be an important one for a better understanding of the singularity at the Cauchy horizon, though we will not address it here.
	
	\section{Geometry inside the mass-inflated region}\label{secri}
	
	This shell-based construction captures (albeit a simple variant of) the mass inflation process and it can give us some additional insights into the behaviour of the geometry inside the black hole at finite times. In particular, to set up our subsequent semiclassical analysis, we want to see how this model answers two questions about the depths of the mass-inflated region. The first one is whether a stream of outgoing radiation below the initial outgoing shockwave may also give rise to an effect similar to mass inflation, which would further increase the rate of growth of mass in the innermost region of the geometry, in the vicinity of the origin. The second one is what particular path the inner apparent horizon actually follows in this whole process, and whether it collapses to the origin to give rise to a spacelike singularity at a finite time $v$, as suggested in the numerical analysis of \cite{Brady1995}.
	
	To answer these questions, we must begin by getting a better understanding of the global structure of the geometry corresponding to the above construction with just a single gravitating outgoing shell. The metrics on the outside and inside of the outgoing shell respectively can be written as
	\begin{align}
		ds^2&=-f(v,r)dv^2+2dvdr+r^2d\Omega^2,\\
		\begin{split}
			dS^2&=-F(v,r)dV^2+2dVdr+r^2d\Omega^2\\&=A(v)\left[-A(v)F(v,r)dv^2+2dvdr\right]+r^2 d\Omega^2,
		\end{split}\label{geo}
	\end{align}
	where lower and upper case letters are once again used for quantities in the regions outside and inside the shell, respectively, and $A(v)=dV/dv=f/F|_{\rm shell}$ is a positive function which allows us to switch between the Eddington-Finkelstein coordinates of these two regions, as expressed on the right-hand side of the latter equation.\par
	For simplicity, we will take a smoothed-out average of the metric functions from the previous section (particularly, their dominant behaviour at late times). The redshift function on the shell evaluated on the inside \eqref{massinfl} is then
	\begin{equation}\label{rb}
		F|_{\rm shell}=-|F_0|e^{\kappa v},
	\end{equation}
	and, from eq. \eqref{ingredient1} and the shell trajectory in each patch, the redshift function on the outside satisfies
	\begin{equation}
		f|_{\rm shell}= -\kappa\left[r_{\rm shell}-r_{\rm i}(v)\right]=-\frac{b}{v^p},
	\end{equation}
	with $b$ a positive constant which depends on the average spacing between ingoing shells (i.e. the average of the quantity $e^{\kappa\Delta v_n}$), into which $\kappa$ (the outside region's inner gravitational radius's surface gravity) and the constant $\beta$ from eq. \eqref{ingredient1} have also been absorbed. The function which relates the outside and inside times, $v$ and $V$, then becomes
	\begin{equation}\label{A1}
		A(v)=\frac{b}{|F_0|}\frac{e^{-\kappa v}}{v^p}.
	\end{equation}
	
	The results we have obtained thus far are valid for any redshift function, but if we want to analyse the deeper regions of the geometry we have to be more specific. Let us therefore first focus on the particular example we have used previously, namely the Reissner-Nordström geometry. If we assume that the ingoing and outgoing shells carry no electrical charge, we can take the variation in $F$ to be solely due to a variation of the mass term. Then, from eqs. \eqref{rb} and \eqref{A1}, and with the Reissner-Nordström redshift function, we get the geometry for the mass-inflated interior region \eqref{geo} written in the $v$ coordinate
	\begin{equation}\label{inflRN}
		\begin{split}
			dS^2=\frac{b}{|F_0|v^p}&e^{-\kappa v}\left[-\frac{b}{v^p}e^{-\kappa v}\left(1-\frac{2M_0e^{\kappa v}}{r}+\right.\right.\\&\left.\left.+\frac{Q^2}{r^2}\right)dv^2+2dvdr\right]+r^2d\Omega^2,
		\end{split}
	\end{equation}
	where $M_0$ is a positive constant which represents the initial mass of the black hole. Using this metric we see that the equation for outgoing null geodesics (which we can later relate to trajectories of additional gravitating outgoing shells) then takes the form
	\begin{equation}\label{outnull}
		\frac{dr}{dv}=\frac{b}{v^p}\left[-h_0(r)+h_1(r)e^{-\kappa v}\right],
	\end{equation}
	where $h_0(r)=M_0/r$ and $h_1(r)=(1+Q^2/r^2)/2$ are positive functions. Due to the exponential suppression of the $h_1$ term, it is clear that the term with $h_0$ is dominant on the right-hand side of this equation, except in a progressively smaller region around the origin, where the inner apparent horizon is shrinking toward zero radius. 
	In relation to this, we will be able to distinguish between two types of outgoing geodesics in the trapped region: ones whose dynamics is predominantly determined by the $h_0$ term, and ones for which the two terms are comparable. The former exist up to $v\to\infty$ only if $p>1$, which, given eq. \eqref{rirn}, is directly related to the mass thrown into the black hole from the outside being finite (for our current example of Reissner-Nordström).\footnote{It is interesting to note that it is quite easy to imagine a geometry in which e.g. due to a relation $\delta r_{\rm i}\propto -(\delta m)^{1/2}$, $p$ ends up being 1 or smaller for a finite accretion of mass. The absence of solutions of the type \eqref{trapped} would then imply a lack of a Cauchy horizon below the outgoing shell, leaving just a trapped region with a tendency toward the formation of a spacelike or null singularity.} If this condition is met and the $h_0$ term continues to dominate, equation \eqref{outnull} generally integrates asymptotically to
	\begin{equation}\label{trapped}
		r=r_{\rm c}+\frac{h_0(r_{\rm c})}{p-1}\frac{b}{v^{p-1}}+\mathcal{O}\left[\left(\frac{b}{v^{p-1}}\right)^2\right].
	\end{equation}
	The integration constant $r_{\rm c}$ represents the final radial position of each light ray when it reaches the Cauchy horizon at $v\to\infty$, which is different for each geodesic depending on initial conditions. In other words, between the outgoing shell and the inner apparent horizon (which is rapidly shrinking toward $r=0$) there are null geodesics which are trapped in a tendency toward a finite radial position which is different from the asymptotic position of this horizon. To see how this journey is perceived from the point of view of an observer in the interior region itself, we can look at the relation between the $v$ and $V$ coordinates, the latter of which, for observers not tending toward the inner apparent horizon, is proportional to the geodesic affine parameter. As we have seen, this relation is given by
	\begin{equation}
		\frac{dv}{dV}=\left.\frac{F}{f}\right|_{\rm shell}=\frac{1}{A(v)},
	\end{equation}
	which asymptotically integrates to
	\begin{equation}\label{trappedtime}
		V=V_{\rm c}-\frac{b}{|F_0|\kappa}\frac{e^{-\kappa v}}{v^p}+\order{\frac{e^{-\kappa v}}{v^{p+1}}}.
	\end{equation}
	These geodesics therefore reach the Cauchy horizon at a finite time parameter
	$V=V_{\rm c}$, corresponding to the integration constant of the equation. This gives us an interpretation for the behaviour seen in \eqref{trapped}: outgoing null geodesics are trapped in a tendency toward these different finite radial positions $r_{\rm c}$ because the function $A$ acts to quickly freeze the proper time, and consequently the movement, of observers in this region. Because of this \textit{freezing function} $A$, most of the outgoing radiation in the trapped region would in fact reach the Cauchy horizon before getting close to the inner apparent horizon. This tells us that interactions between the ingoing shells with additional outgoing shells travelling along these geodesics would not have time to produce any sort of amplification of the mass inflation effect, as this would require proximity to the inner gravitational radius $R_{\rm i}$ of this internal region (which in the absence of such shells is in fact the inner apparent horizon position).
	
	The region where eqs. \eqref{trapped} and \eqref{trappedtime} are valid begins on the inside of the outgoing shell and increases in size toward the origin as the mass tends to infinity due to its unbounded growth following eqs.~\eqref{Misner-Sharp} and~\eqref{rb}. The radii $r_{\rm c}$ at which null geodesics freeze can then vary continuously from $r_{\rm i}$ (the inner gravitational radius of the external geometry) to zero, as is observed in the analytical study of the Cauchy horizon in \cite{Ori1998}. However, this does not imply that all outgoing null geodesics are trapped in this way and are unaffected by the shrinking inner apparent horizon. Due to the growing mass, the radial position of this apparent horizon $R_{\rm i}$ can be approximated by a series expansion of the lower of the two roots of the Reissner-Nordström redshift function in $1/M(v)$, revealing its tendency to zero
	\begin{equation}\label{ria}
		R_{\rm i}(v)=\frac{Q^2}{2M(v)}+\order{\frac{1}{M(v)}}^3=\frac{Q^2}{2M_0}e^{-\kappa v}+\order{e^{-3\kappa v}},
	\end{equation}
	with a surface gravity (in absolute value)
	\begin{equation}\label{RNk1}
		K_{\rm i}(v)=\frac{b}{v^p}\frac{2M_0^3}{Q^4}e^{2\kappa v}+\order{\frac{1}{v^p}}.
	\end{equation}
	There are indeed many outgoing null geodesics which are sufficiently close to this horizon to approach it asymptotically, i.e. to tend to zero from above it. These are the second-type geodesics we mentioned before: the ones for which the $h_0$ and $h_1$ terms of eq. \eqref{outnull} are comparable, with the right-hand side of this equation being close to zero, i.e. $r$ being close to $R_{\rm i}$. Their movement can be described by expanding the right-hand side of this equation around $R_{\rm i}$,
	\begin{equation}\label{geo-ria}
		\frac{dr}{dv}\simeq-K_{\rm i}(v)[r-R_{\rm i}(v)].
	\end{equation}
	With the rapidly growing value of $K_{\rm i}$, it can be readily checked that this equation has a family of solutions for which $r\to R_{\rm i}$. These geodesics also reach $v=\infty$ with a finite affine parameter, where they converge to $r=0$ along with $R_{\rm i}$, falling into a (strong) curvature singularity.
	
	If an additional mass inflation effect can take place, pushing the position of the inner apparent horizon to below the one given by~\eqref{ria}, it would be triggered by placing an outgoing shell precisely on one of the geodesics described by~\eqref{geo-ria} which tend toward $R_{\rm i}$ from above. To keep the label $R_{\rm i}$ for the zero of the $F$ function~\eqref{ria}, i.e. the inner gravitational radius of the $F$ patch, we will now call the actual inner apparent horizon $\tilde{R}_{\rm i}$. Also, instead of working with the $v$ coordinate, here it will be more convenient to use $V$, with which we can directly apply the relation \eqref{quotient} for the junction conditions on the intersection points with the ingoing shells. The position of the inner gravitational radius $R_{\rm i}$ and its surface gravity (taken as the absolute value of $\partial_rg_{vv}/(2g_{vr})$ in each coordinate system)\footnote{In other words, $K_{\rm i}^V$ is not just $K_{\rm i}$ with a coordinate change, due to the $A(v)$ factor which multiplies the redshift function after the coordinate change, as can be seen in \eqref{inflRN}.} in the $V$ coordinate system $K_{\rm i}^V$ evolve as
	\begin{align}
		R_{\rm i}&=\xi_1(V_{\rm c}-V)+\mathcal{O}[(V_{\rm c}-V)^3],\\
		K_{\rm i}^V&=\frac{\xi_2}{(V_{\rm c}-V)^3}+\order{\frac{1}{V_{\rm c}-V}}
	\end{align}
	where $\xi_1$ and $\xi_2$ are some positive constants which depend of the asymptotic mass, the charge and the initial conditions. In the shell model we once again consider that these functions actually have discrete jumps at a set of points in $V$, corresponding to the positions of the infalling shells. If we take these shells to be either equispaced in $v$, or at most distributed with a density polynomially dependent in $v$, then in $V$ their spacing decreases as
	\begin{equation}
		\Delta V_n=V_n-V_{n-1}\sim V_{\rm c}-V_n
	\end{equation}
	as $V_n$ tends toward the Cauchy horizon $V_{\rm c}$. The jumps in the position of the inner horizon between shells also decrease as
	\begin{equation}
		\Delta R_{\rm i,n}\sim V_{\rm c}-V_n.
	\end{equation}
	Eq. \eqref{quotient} can be applied directly here because the outgoing shell has enough time between each iteration to get exponentially closer to $R_{\rm i,n}$, as can be seen by the fact that
	\begin{equation*}
		e^{-K^V_{\rm i,n}\Delta V_{n}}\sim e^{-1/(V_{\rm c}-V_{n})^2}\to 0, \quad \text{as}\quad V_{n}\to V_{\rm c}.
	\end{equation*}
	The quotient of surface gravities in \eqref{quotient} once again tends to a constant in the limit of interest, and so does the quotient of the differences between radial positions in front of the exponential.\footnote{This can be seen explicitly by solving the outgoing null geodesic equation between each infalling shell and matching the solutions. The calculation is completely analogous to the one performed in the vicinity of $r_{\rm i}$.} This leaves the redshift function below this new outgoing shell with an increase given by a multiplicative factor
	\begin{equation*}
		e^{K^V_{\rm i,n}\Delta V_{n}}
	\end{equation*}
	after each iteration, which diverges as $e^{\xi_2/(V_{\rm c}-V)^2}$ toward the Cauchy horizon. In terms of $v$, this new mass inflation increases the Misner-Sharp mass in the vicinity of the origin as an \textit{exponential of an exponential}. One can then imagine that each time we repeat this whole process considering the new displacement of the inner horizon caused by this effect, we would get an additional exponential in $v$ to the chain.
	
\begin{figure}
		\centering
		\includegraphics[scale=1.2]{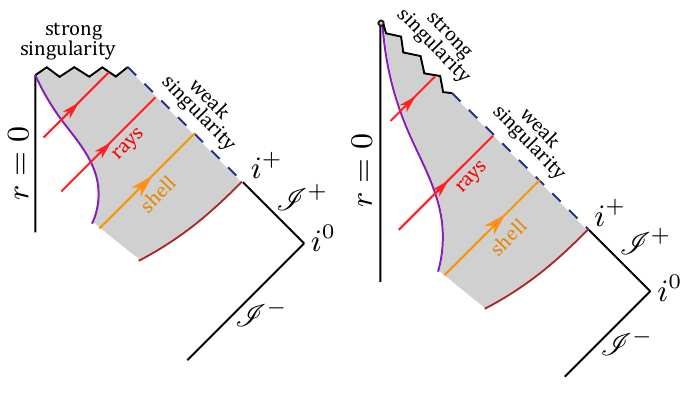}
		\caption{Future part of the causal diagram of the mass inflation geometry. The shaded part is the trapped region and the dashed line is the Cauchy horizon (and a weak singularity). The null shell shown is the upper outgoing one which tends to $r_{\rm i}$ and is responsible for the first mass inflation effect. Left: the inner apparent horizon reaches $r=0$ at finite $v$ and forms a Schwarzschild-type spacelike singularity. Right: the inner horizon only tends to $r=0$ asymptotically in $v$, resulting in a strong null singularity at $v\to\infty$ and $r=0$, at a finite affine distance for geodesics which fall into it.}
		\label{f3}
	\end{figure}

	We can now tackle answering the second question posed at the beginning of this section: what is the actual path followed by the inner horizon, and what causal structure does this movement give rise to. It is hardly surprising that in the case of continuous matter the chain effect just described could produce a spacelike singularity at finite $v$, as is observed numerically in \cite{Brady1995} and commented on in later works \cite{Ori1998,Marolf2012}. Even if the exact analytical solution were to only result in a very quick tendency toward the formation of this singularity (i.e. a quick, but still asymptotic in $v$, approach of the inner apparent horizon toward the origin), this approach would in fact be so quick that it would likely be numerically indistinguishable from a singularity at a finite time $v$; at any rate, the curvature would become Planckian very fast, making a classical description of the geometry-matter interactions inadequate. However, it is interesting to note that within the classical description, this seemingly small difference between a singularity forming at strictly finite $v$ and having just a tendency toward its formation, however quick, and only forming it at $v\to\infty$, results in two different asymptotic structures, represented in the two diagrams of fig.~\ref{f3}. In fact, the former case results in the formation of a Schwarzschild-type spacelike singularity, while the latter ends in a (strong) null singularity, both of which are at $r=0$ and are attached to the weak null singularity which spans the Cauchy horizon, where $r$ takes values up to $r_{\rm i}$. To clarify, we use the same criterion for distinguishing strong and weak singularities as the one described in~\cite{Ori1991,EllisSchmidt1977}: both have diverging curvature scalars, but the distortion suffered by an observer of finite size remains bounded when crossing a weak singularity, while it diverges when crossing a strong one. One can readily check that the curvature blow-up of the geometry we use close to the Cauchy horizon [see eq. \eqref{inflRN}] is the same as in refs.~\cite{Ori1991,Brady1995}, $R_{\mu\nu\rho\sigma}R^{\mu\nu\rho\sigma}\sim 48M_0^2r^{-6}e^{2\kappa v}$, leading to the same type of weak singularity there.
	
	Up to here, all we have said regarding these questions has been based on the Reissner-Nordström background. For other geometries besides Reissner-Nordström, the inner structure of the mass-inflated region depends on how the geometry reacts to the increase in mass provided by the infalling null shells, particularly on the trajectory followed by the inner gravitational radius. The relation from the charged black hole $\delta r_{\rm i}\propto(-\delta m)$ is also satisfied in the case of a rotating black hole, but in regular black hole spacetimes this tendency may be modified, depending on how the regularisation of the origin is achieved in the first place. One may expect there to exist trapped geodesics of the type~\eqref{trapped}, but the structure toward the origin may differ, possibly avoiding the formation of a strong singularity altogether by preventing the inner horizon from getting too close to $r=0$.
	
	It is also interesting to note that in general, the explicit divergence of the mass at $v\to\infty$ depends on the ever-smaller ingoing perturbations also continuing up to infinity. Classically it is perfectly natural to consider this to be the case, but one may imagine that a quantum description of the interaction between the black hole and infalling matter may have a lower bound on the energy which can actually affect the black hole. For example, we can consider that this lower bound is given by the energy of a single photon with a wavelength of the order of the black hole external mass $m$ (in geometric units). Then, we can relate this energy to a mass and to a cutoff time $v_{\rm cut}$ through \eqref{deltam}, and for simplicity we can take the polynomial tail in \eqref{ingredient1} to be the same as in \eqref{deltam}, i.e. $p=q$. The result is that by $v_{\rm cut}$ the mass in the interior region would have increased by a factor $e^{(M/l_{\rm P})^{2/p}}$ from just the first exponential mass inflation effect around $r_{\rm i}$, given by eqs. \eqref{massinfl} and \eqref{Misner-Sharp}. Needless to say, the exponent in this number is generally very large. For a solar mass black hole and a polynomial decay with $p=12$ (as considered in \cite{PoissonIsrael89}) the mass grows by a factor larger than $e^{10^{6}}$, making the dynamics of the interior region enter a full quantum gravity regime.
	
	In conclusion, we can say that classically mass inflation would occur under quite generic circumstances for charged black holes, and by extension for rotating ones as well, where an analogous shell-based construction can be made \cite{Barrabes1990}. While in principle it is possible for the perturbations falling in to carry enough charge or angular momentum to keep the inner horizon still enough for the instability to not be triggered, this is unlikely to occur in an astrophysical scenario over long periods of time. The only viable way to avoid the instability seems to be to rely on a regularising mechanism which prevents the inner horizon from getting too close to the origin. We will now explore whether the quantum nature of matter can provide such a regularising tendency in the semiclassical gravity regime.
	
	\section{Semiclassical backreaction}\label{secsemicl}
	
	The main goal of this work is to see whether and how semiclassical physics can have an influence on the evolution of the inner horizon inside a black hole undergoing mass inflation. A recent work by the present authors \cite{Barcelo2021} analyses the backreaction around static and dynamical inner horizons in simple geometries which do not incorporate mass inflation. The result was that the inner horizon tends to be pushed outward due to semiclassical effects. The initial tendency for its movement is exponential in time, and an extrapolation of the process can thus be dubbed an \emph{inflation of the inner horizon}, which leads to a quick extinguishing of the trapped region from the inside out. In this section we will extend our semiclassical backreaction analysis to geometries incorporating the mass inflation effect, the main new component being the presence of the freezing function $A(v)$.
	
\begin{figure}
		\centering
		\includegraphics[scale=.7]{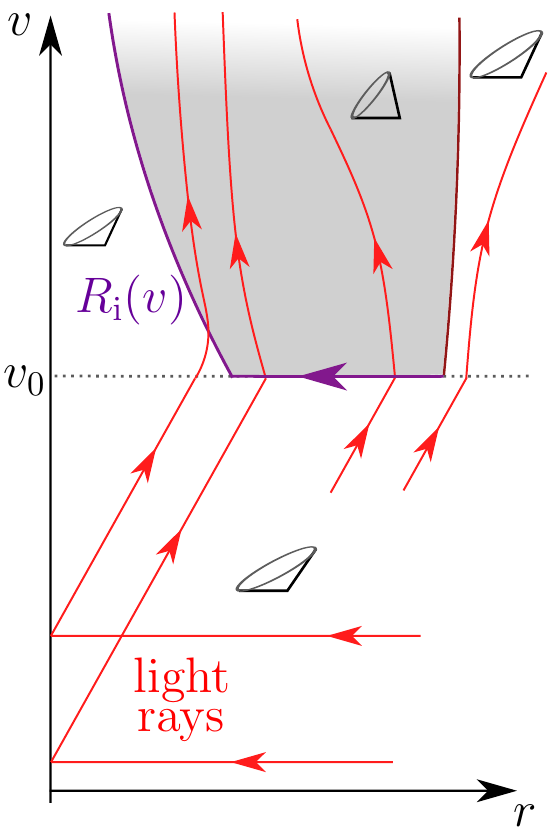}
		\caption{Formation of a black hole by a ingoing null shell at $v=v_0$. The ``in" vacuum state is constructed by tracing light rays back to the flat region in the past. Classically, the black hole is undergoing mass inflation and the inner horizon is headed toward the origin.}
		\label{fin}
	\end{figure}

	As in \cite{Barcelo2021}, we will use the RSET of a massless scalar field in the Polyakov approximation as the source of backreaction. Although this approximation is far from being the exact RSET of a 3+1 dimensional geometry, it is sufficient to describe Hawking evaporation at the outer horizon, and is therefore a good candidate to give us a first glimpse at horizon-related semiclassical effects around the inner apparent horizon. We construct the ``in" vacuum \cite{Fabbri2005,Frolov2017} of the 1+1 dimensional radial-temporal sector of the spacetime, used to calculate the RSET in this approximation, by following the movement of light-like geodesics from past null infinity up to the region of interest, which in this case is the vicinity of the inner apparent horizon inside the mass-inflated region of a dynamically formed black hole. As discussed in \cite{Barcelo2021}, the accumulative effect that the inner horizon has on null geodesics results in a sensitivity of the RSET to the past of a large part of the collapse geometry, unlike what occurs for the outer horizon. We therefore employ the same tactic as in that work to simplify the initial conditions of the quantum modes in this region: we consider their propagation as being in a flat spacetime up to an advanced time $v=v_0$, from where it continues inside a black hole with an inner horizon, as shown in fig. \ref{fin}. Effectively this is equivalent to the black hole being generated by the collapse of a null shell located at the formation time $v=v_0$, but our motivation for the construction is purely geometrical, to ``clean up" the dependence on the details of the collapse and leave only the part stemming from the quantum modes finding themselves in the trapped region.
	
	In this section, we will perform two distinct calculations to estimate semiclassical backreaction in the vicinity of the inner horizon. The first involves a series expansion of the RSET and the metric functions around the point $\{v_0,R_{\rm i0}\}$, where the inner horizon $R_{\rm i}$ forms, which allows a simplified term-by-term calculation of the semiclassical perturbations of the metric caused by the RSET. The second is a full self-consistent solution for a classical background with a particular family of functions $F(v,r)$, valid for a small but finite time interval after $v_0$. From the latter calculation we find that the initial tendencies seen in the series expansion can lead to very quick accumulative effects which make semiclassical corrections relevant before the spacetime curvature reaches Planckian scales.
	
	\subsection{Series expansion around horizon formation}
	
	Let us begin by considering the line element
	\begin{equation}\label{geoinfl}
		ds^2=A(v)\left[-B(v,r)dv^2+2dvdr\right]+r^2d\Omega^2.
	\end{equation}
	We will use this type of geometry to represent the inner region of a black hole undergoing mass inflation, particularly around its inner apparent horizon. In relation to the previous section, we are using the geometry obtained with a single outgoing shell, resulting in a single exponential growth of mass, which we expect to be the dominant effect in a transient period between early and very late times, where we place our $v_0$. For the freezing function $A$ we can use the dominant behaviour in $v$ of~\eqref{A1} and set $A(v)=e^{-\kappa v}$, with $\kappa$ a positive constant.\footnote{Neglecting the $1/v^p$ part of \eqref{A1} amounts to discarding corrections suppressed by an additional $1/v$ factor in the backreaction calculations of this section, which do not have a qualitative influence on our conclusions.} For the function $B$, which represents the product of $F$ and $A$ from~\eqref{geo}, we only need to impose that it has a zero at a radius $R_{\rm i}(v)$, corresponding to the inner apparent horizon, with a negative slope in the $\partial_r$ direction.
	
	To calculate the RSET for this spacetime, we first need to construct a quantisation with a vacuum state which is physically adequate for the problem at hand. We will use the ``in" vacuum state, which is defined as the Minkowski vacuum at the asymptotically flat region of past infinity and its extension to the dynamical region through the propagation of the particle-related modes. In the Polyakov approximation, we only need to do this in the 1+1 dimensional radial-temporal sector of our geometry, which simplifies the problem greatly, given that 1+1 dimensional spacetimes are conformally flat. Working with a pair of radial null coordinates $\{u,v\}$, the line element \eqref{geoinfl} can generally be written as
	\begin{equation}
		ds^2=-C(u,v)dudv+r^2d\Omega^2.
	\end{equation}
	At past null infinity, $C\sim 1$ and Poincaré invariance of the vacuum state in flat spacetime amounts to supertranslation invariance of the ``in" state \cite{Hawking1975}. The $v$ coordinate used in \eqref{geoinfl} is in fact one of these ``in" coordinates, while $u$ must be obtained through its relation to $r$.
	
	The conformal factor of the radial-temporal sector is given by
	\begin{equation}\label{conformal}
		C(u,v)=-2A(v)\frac{\partial r(u,v)}{\partial u}.
	\end{equation}
	The components of the RSET in the Polyakov approximation for the vacuum state selected by the coordinate system $\{u,v\}$ are \cite{Fabbri2005,Barcelo2012}
	\begin{subequations}\label{RSETg}
		\begin{align}
			\expval{T_{uu}}&=\frac{l_{\rm P}^2}{96\pi^2r^2}\left[\frac{\partial_u^2C}{C}-\frac{3}{2}\left(\frac{\partial_uC}{C}\right)^2\right],\\
			\expval{T_{vv}}&=\frac{l_{\rm P}^2}{96\pi^2r^2}\left[\frac{\partial_v^2C}{C}-\frac{3}{2}\left(\frac{\partial_vC}{C}\right)^2\right],\\
			\expval{T_{uv}}&=\frac{l_{\rm P}^2}{96\pi^2r^2}\left[\frac{\partial_uC\partial_vC}{C^2}-\frac{\partial_u\partial_vC}{C}\right],
		\end{align}
	\end{subequations}
	where $l_{\rm P}$ is the Planck length. This tensor has a non-physical divergence at the origin due to the process of dimensional reduction, making it inadequate to use as such for the backreaction problem in the whole spacetime. However, as discussed earlier, it is an adequate probe of quantum effects in the vicinity of horizons, where it captures the non-local behaviour which e.g. leads to the Hawking effect \cite{DFU}. In this case, we will only use it in a small vicinity around the inner horizon, and only while this horizon is considerably farther away from the origin than a Planck length (i.e. early enough in the evolution of the mass inflation background).
	
	To calculate the conformal factor \eqref{conformal}, we need to obtain the function $r(u,v)$ from the solutions of radial null geodesics. The ingoing geodesics are just $v=\text{const.}$, while the outgoing ones are solutions to
	\begin{equation}\label{geodesic}
		\frac{dr}{dv}=\frac{1}{2}B(v,r).
	\end{equation}
	
	At this point we must either specify the function $B$, or try to see what general conclusions could be obtained from just the mere fact that there is an inner horizon in this structure, i.e. that $B$ has a zero at some $R_{\rm i}(v)$ with a negative slope. In the next subsection we will specify some functions $B$ which can simplify our analysis while still reproducing the causal properties of mass inflation, but for now we will maintain generality and perform a perturbative analysis. Particularly, we will consider a generic expansion of the function $B$ around the point at which this horizon forms $\{v_0,R_{\rm i0}\}$,
	\begin{align}
		\begin{split}
			\frac{1}{2}B(v,r)&=k_1(v)(r-R_{\rm i}(v))+k_2(v)(r-R_{\rm i}(v))^2+\\&\quad+k_3(v)(r-R_{\rm i}(v))^3+\cdots,\end{split}\label{redshiftseries}\\
		R_{\rm i}(v)&=R_{\rm i0}+R_{\rm i1}v+R_{\rm i2}v^2+R_{\rm i3}v^3+\cdots,\\
		k_n(v)&=k_{n0}+k_{n1}v+k_{n2}v^2+k_{n3}v^3+\cdots,\label{kn}
	\end{align}
	with $n=1,2,\dots$, and where for simplicity we have set $v_0=0$. The only conditions we impose on these series is that $R_{\rm i0}>0$ and $k_{10}<0$ (this being the inner horizon). The smallness of the terms in the expansions of quantities with (inverse) length dimensions, here and throughout this section, can be measured in terms of their respective initial values at $v=0$, or in units of the characteristic initial scale $R_{\rm i0}$.
	
	For the solution of \eqref{geodesic} we consider the series expansion
	\begin{equation}\label{null}
		r(v)=r_0+r_1v+r_2v^2+r_3v^3+\cdots
	\end{equation}
	Substituting this expression into \eqref{geodesic}, we obtain the coefficients of the null trajectories \eqref{null} in terms of derivatives of $B$ [i.e. the coefficients of its expansion \eqref{redshiftseries}-\eqref{kn}] and a free parameter fixed by an initial condition. We will use $d_0=r_0-R_{\rm i0}$ as this parameter. Tracing back the null trajectories through the Minkowski region $v<0$ (see fig.~\ref{fin}) we find that our missing ``in" coordinate is $u=-2d_0$ (up to a constant which fixes the origin of $u$, taken as zero).
	
	Constructing $r(u,v)$ in this manner, we calculate the conformal factor \eqref{conformal} and then the RSET components~\eqref{RSETg} in the ``in" coordinate system. Switching them back to the Eddington-Finkelstein system, at zeroth order in the series expansion, they are
	\begin{subequations}
		\begin{align}
			\begin{split}
				\expval{T_{vv}}&=\frac{l_{\rm P}^2}{96\pi^2R_{\rm i0}^2}\left(-\frac{1}{2}k_{10}^2+k_{11}-2k_{20}R_{\rm i1}-\right.\\&\qquad\qquad\qquad\left.-\frac{1}{2}\kappa^2+k_{10}\kappa\right)+\order{v,d},\end{split}\\
			\expval{T_{rr}}&=\order{v,d},\\
			\expval{T_{vr}}&=-\frac{2l_{\rm P}^2k_{20}}{96\pi^2R_{\rm i0}^2}+\order{v,d},
		\end{align}
	\end{subequations}
	where $d=r-R_{\rm i0}$ [here $\order{d}$ can be though of as $\order{d_0}$ or $\order{u}$, as the difference between them is $\order{v}$].
	
	We now consider the semiclassical Einstein equations
	\begin{equation}\label{semicl}
		G_{\mu\nu}+\delta G_{\mu\nu}=T_{\mu\nu}^{\rm class}+\expval{T_{\mu\nu}},
	\end{equation}
	where $\delta G_{\mu\nu}$ is a perturbation to the background Einstein tensor and $T_{\mu\nu}^{\rm class}$ is the classical matter content sourcing the zeroth order background. In general, the perturbation caused by the RSET would also affect $T_{\mu\nu}^{\rm class}$ through its dependence on the metric. However, since we want to remain as agnostic as possible about this classical matter, we consider that $\delta T_{\mu\nu}^{\rm class}=0$ at zeroth order in the series expansion (in its functional form in Eddington-Finkelstein coordinates), i.e. that
	\begin{equation}\label{einsteinpert}
		\delta G_{\mu\nu}=\expval{T_{\mu\nu}}+\order{v,d}.
	\end{equation}
	This simplifying assumption serves two purposes: on the one hand, it allows us to continue to work in purely geometric terms, with as few ingredients in the dynamics as possible. On the other hand, it exemplifies well what backreaction from the RSET can look like in its purest form, where the potentially negative-energy terms in this tensor directly source $\delta G_{\mu\nu}$. Technically, considering $\delta T_{\mu\nu}^{\rm class}=0$ would over-determine the system of equations, but it turns out to work consistently up to second order in our series expansion, allowing us this first geometric glimpse into backreaction.
	
	Using the expansion of the function $B$ and its coefficients, the Einstein tensor of our generic background is
	\begin{subequations}\label{einstein}
		\begin{align}
			G_{vv}&=\frac{2k_{10}R_{\rm i1}}{R_{\rm i0}}+\order{v,d},\\
			G_{rr}&=\order{v,d},\\
			G_{vr}&=\frac{2k_{10}R_{\rm i0}-1}{R_{\rm i0}^2}+\order{v,d}\\
			G_{\theta\theta}&=2R_{\rm i0}(k_{10}+k_{20}R_{\rm i 0}).
		\end{align}
	\end{subequations}
	To construct $\delta G_{\mu\nu}$ we can consider perturbations to these coefficients, e.g. $k_{10}\to k_{10}+\delta k_{10}$. Eq. \eqref{einsteinpert} allows us to fix one of the three coefficients present in the tensor components \eqref{einstein} to its classical value as an initial condition, and we choose the initial position of the inner horizon $R_{\rm i0}$. Then, perturbing the surface gravity $k_{10}$ and the initial time derivative of the inner horizon trajectory $R_{\rm i1}$, and using \eqref{einsteinpert} we obtain two key relations,
	\begin{align}
		\delta k_{10}&=-\frac{l_{\rm P}^2}{12\pi}\frac{k_{20}}{R_{\rm i0}},\\
		\delta R_{\rm i1}&=\frac{l_{\rm P}^2}{48\pi}\left(-\frac{k_{10}}{R_{\rm i0}}-\frac{\kappa^2}{R_{\rm i0}k_{10}}+\frac{1}{2}\frac{\kappa}{R_{\rm i0}}+\frac{2k_{11}}{R_{\rm i0}k_{10}}\right).\label{dr}
	\end{align}
	On the one hand, we can see that the semiclassical contribution to the modification of the surface gravity can be either positive or negative, depending on the sign of the background coefficient $k_{20}$. This initial semiclassical contribution very much depends on the details of the initial background geometry. On the other hand, the modification of the derivative of the inner horizon trajectory is almost always positive, implying a decrease in the rate at which it moves inward. This is a first indication of the regularising tendency which semiclassical corrections can add to the inner horizon dynamics in these spacetimes. Particularly, it can be seen from the fact that $k_{10}<0$, $\kappa>0$, $R_{\rm i0}>0$, and the assumption that $k_{11}<0$, i.e. that the background surface gravity tends to become increasingly more negative, which is certainly the case in mass inflation, as can be seen from e.g. eq. \eqref{RNk1}. Interestingly, the further along an evolution of the type \eqref{ria} we set our initial conditions for semiclassical backreaction, the smaller the background value of $R_{\rm i1}$ would be (approaching zero as $v\to\infty$) and the larger $\delta R_{\rm i1}$ would be in comparison. If the background surface gravity  $k_{10}$ has a divergent behaviour akin to \eqref{RNk1}, then the growth of $\delta R_{\rm i1}$ as we take our initial radial position $R_{\rm i0}$ to 0 cannot be said to be a consequence of the unphysical divergent $1/r^2$ factor present in the RSET approximation. More generally, if $k_{10}$ diverges at least as strongly as $1/R_{\rm i0}$, then a regularised version of this perturbation $R_{\rm i0}^2\delta R_{\rm i1}$ remains at least finite, as opposed to the background $R_{\rm i1}$ which is expected to approach zero unless a spacelike singularity forms at finite $v$.
	
	Using these initial tendencies as an estimate of the magnitude of this effect later on in the evolution, one is led to the hypothesis that semiclassical backreaction will always become dominant at some point before a singularity is formed, perhaps leading instead to a non-singular future. The only caveat might appear when interpreting the semiclassical effects as only the first set of corrections towards a quantum gravity theory. Then, if the corrections occurred only when curvature becomes Planckian, i.e. when the background values in \eqref{dr} overcome the suppression by the $l_{\rm P}^2$ factor, one could argue that these semiclassical effects would have been already superseded by other effects of unknown character. However, as we will see in the following example, a full time evolution can lead to a very different result, in which semiclassical corrections have a much quicker accumulative effect.
	
	\subsection{Time-integrable example}
	
	In order to see what the semiclassical evolution of the inner horizon could look like beyond the initial tendencies calculated above, we can use a particular family of geometries for the classical background which simplify our semiclassical analysis greatly. Particularly, we will use geometries which, in a vicinity around the inner horizon, take the form \eqref{geoinfl} with
	\begin{equation}\label{redshiftcl}
		B(v,r)=e^{-\kappa v}-\frac{1}{2}\lambda(v)r,
	\end{equation}
	where $\lambda(v)$ is a positive, but otherwise arbitrary function. The inner horizon described by this geometry,
	\begin{equation}
		R_{\rm i}(v)=2\frac{e^{-\kappa v}}{\lambda(v)},
	\end{equation}
	moves toward the origin as long as $\lambda$ does not decrease faster than $e^{-\kappa v}$. The relation between this position and its surface gravity is not quite the same as in e.g. the Reissner-Nordström case seen above (where for $R_{\rm i}\sim e^{-\kappa v}$, the surface gravity has an increase with a rate $e^{2\kappa v}$), but a growing surface gravity can still be replicated by choosing a $\lambda$ which increases in time.
	
	The RSET in the Polyakov approximation corresponding to this classical background geometry has a single non-zero component: the ingoing flux
	\begin{equation}\label{RSET}
		\expval{T_{vv}}=\frac{l_{\rm P}^2}{96\pi^2r^2}\left[-\frac{1}{4}\lambda'-\frac{1}{32}\lambda^2-\frac{\kappa}{4}\lambda-\frac{\kappa^2}{2}\right],
	\end{equation}
	which is negative as long as $\lambda'\geq0$ (this being a reasonable requirement for a mass inflation background, i.e. that the surface gravity of the inner horizon does not decrease).
	
	The main motivation for using these geometries is that, as we will see, backreaction from the RSET around the inner horizon has the effect of changing the $B$ function to
	\begin{equation}\label{redshift}
		B(v,r)=e^{-\kappa v}-\frac{1}{2}\lambda(v)r+\delta B(v,r),
	\end{equation}
	where the first two terms are just its background form, and the perturbation $\delta B$, obtained from the semiclassical Einstein equations, will have a particularly simple form in terms of its dependence in $r$,
	\begin{equation}\label{Br}
		\delta B(v,r)=-\frac{\alpha(v)}{r}
	\end{equation}
	(the minus sign serves to make an analogy with a mass term, as discussed below). This can be readily checked by calculating the Einstein tensor with \eqref{redshift}, which has the non-zero components
	\begin{subequations}
		\begin{align}
			G_{vv}&=\frac{\lambda'}{2}-\frac{\lambda^2}{2}+\frac{e^{-\kappa v}\lambda}{r}+\frac{\kappa\lambda}{2}+\frac{1}{r^2}\left[\alpha'-(\lambda-\kappa)\alpha\right],\label{Gvv}\\
			G_{vr}&=-\frac{\lambda}{r},\\
			G_{\theta\theta}&=\frac{G_{\phi\phi}}{\sin^2\theta}=-\frac{r\lambda e^{\kappa v}}{2},
		\end{align}
	\end{subequations}
	where primes denote derivatives with respect to $v$. We see that $\alpha$ appears only in $G_{vv}$, and that the terms which contain it can be directly equated to the RSET flux \eqref{RSET}, given that they have the same dependence in $r$ [the form of $\delta B$ in \eqref{Br} was obtained by requiring this]. Here we are once again assuming that the classical part of the equations, i.e. the rest of the terms of $G_{\mu\nu}$ and their source $T_{\mu\nu}^{\rm class}$, remain unchanged. Analogously to the series expansion calculation above, our motivation for doing this is to introduce as little information about the classical matter content as possible, while also getting a cleaner backreaction problem for the Einstein tensor sourced solely by the RSET (in this case it is the particular form of the background which allows us to do this without over-determining the system of equations). For as long as this perturbation of non-zero $\alpha$ is negligible for the calculation of the RSET itself, and while \eqref{RSET} is accurate (which is the case for a finite time interval after $v_0$, which is smaller for faster dynamics of the background), this gives an approximate self-consistent solution to the semiclassical Einstein equations.
	
	Note that the function $\alpha$ in $B$ is analogous to the $M_0$ term in \eqref{inflRN}, which is constant in the product of $A$ and $F$ used to construct $B$ (up to the decaying inverse polynomial $1/v^p$ terms, which we have neglected in our construction here), and represents the classically exponentially growing mass. Thus, if $\alpha$ grows, toward either positive of negative values, it could be taken as an indication of the semiclassical backreaction tending to become the dominant source of dynamics. Indeed, we will see that $\alpha$ becomes negative and in many cases its absolute value tends to grow exponentially quickly.
	
	Equating \eqref{RSET} to the terms containing $\alpha$ in \eqref{Gvv}, leaving the remaining terms as fixed by the background, the evolution of this semiclassically sourced $\alpha$ is given by the equation
	\begin{equation}\label{self-consistent}
		\alpha'(v)-\eta_1(v)\alpha(v)=\frac{l_{\rm P}^2}{48\pi}\eta_2(v)+\order{l_{\rm P}^4},
	\end{equation}
	where
	\begin{equation}\label{eta2}
		\eta_1=\lambda-\kappa,\quad \eta_2=-\lambda'-\frac{1}{8}\lambda^2-\kappa\lambda-2\kappa^2
	\end{equation}
	are two functions determined by the choice of background. The general solution of this equation is
	\begin{equation}
		\alpha(v)=\frac{l_{\rm P}^2}{48\pi}e^{\int^v\eta_1(\tilde{v})d\tilde{v}}\left[c_1+\int^ve^{-\int^{\tilde{v}}\eta_1(\bar{v})d\bar{v}}\eta_2(\tilde{v})d\tilde{v}\right],
	\end{equation}
	where $c_1$ is the integration constant which can be fixed by initial conditions.
	
	Let us first look at the simple case in which $\lambda$ is constant, which represents an inner horizon shrinking in radius proportionally to $e^{-\kappa v}$ while maintaining a constant surface gravity. Here we already see the main difference from the situations studied in \cite{Barcelo2021}. Depending on whether $\kappa$ or $\lambda$ is larger, $\eta_1$ can be either a positive or negative constant, while $\eta_2$ is always a negative constant. With the condition of the semiclassical perturbation being initially zero, $\alpha(0)=0$, we get the solution
	\begin{equation}
		\alpha(v)=\frac{l_{\rm P}^2}{48\pi}\frac{\eta_2}{\eta_1}[e^{\eta_1v}-1].
	\end{equation}
	Whether $\eta_1$ is positive or negative, $\alpha$ evolves toward negative values. In the former case its absolute value grows exponentially quickly, while in the latter it tends to a constant. There is also the particular case in which $\lambda=\kappa$, for which, since $\eta_1=0$, the solution is
	\begin{equation}
		\alpha(v)=\frac{l_{\rm P}^2}{48\pi}\eta_2v,
	\end{equation}
	with $\eta_2$ being once again negative. In all these cases, the (increasingly) negative values of $\alpha$ tend to push the inner horizon outward. Expanding the radial position of this horizon around $\alpha=0$,
	\begin{equation}
		R_{\rm i}=2\frac{e^{-\kappa v}}{\lambda}-\alpha e^{\kappa v}+\cdots,
	\end{equation}
	we see that once $\alpha$ acquires a non-vanishing value, even in the case where it tends to a constant, this radius quickly acquires non-perturbative corrections. The inner horizon thus begins to move outward, which, although in apparent violation of causality, is hardly surprising considering that the source given by the RSET \eqref{RSET} is an ingoing flux of negative energy.
	
	For more general backgrounds given by different functions $\lambda(v)$, we can see from equations \eqref{self-consistent} and \eqref{eta2} that starting from $\alpha(0)=0$, $\alpha(v)$ will tend to decrease and the inner horizon will tend to move outward unless $\lambda(v)$ decreases sufficiently quickly for $\eta_2$ to become positive. For example, if $\lambda$ tends to zero asymptotically in $v$ and its tendency is quicker than $1/v$ (but slower than $e^{-\kappa v}$, so the inner horizon does not move outward classically), and if $\kappa$ is initially negligible in \eqref{eta2}, there can be a period of time in which $\alpha$ increases toward positive values. However, except in these specific scenarios, $\eta_2$ will generally be negative and $\alpha$ will acquire negative values, making the semiclassical movement of the inner horizon an outward one.
	
	Therefore, the results obtained in \cite{Barcelo2021} appear to still hold in most of these mass inflation geometries. In other words, while the flux \eqref{RSET} dominates the RSET, backreaction tends to push the inner horizon outward. However, it is worth reminding the reader that the conditions for which we have been able to show that this flux is dominant only hold true for a short period of time after the formation of the black hole, given by the time it takes for outgoing null geodesics which come from outside the region where \eqref{redshift} is accurate to intersect the inner horizon. The result for the movement of the inner horizon is therefore only accurate as an initial tendency. Still, the fact that the RSET is likely to keep violating energy positivity conditions even in the later parts of the evolution makes the possibility that the trapped region continue to evaporate from the inside a likely one.
	
\begin{figure}
		\centering
		\includegraphics[scale=.7]{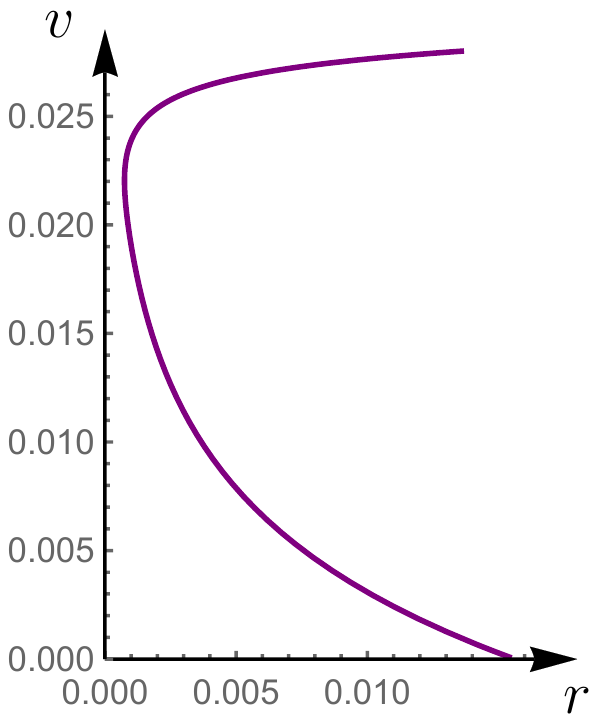}
		\caption{Trajectory of the inner apparent horizon in an extrapolated semiclassical solution, with a classical background which simulates the behaviour of the interior of a mass-inflated charged black hole. Length units are taken in terms of the exterior mass $M_0$, the charge is taken to be $Q=M_0/2$ and the Planck length is set as $10^{-5}M_0$ (as a large difference in orders of magnitude is necessary, but a smaller value only increases the computational difficulty while giving no qualitative difference). Note that the bounce occurs at $r\simeq 10^{-3}M_0$, which illustrates that semiclassical effects can become dominant without the radius of the inner horizon becoming Planckian}.
		\label{f4}
	\end{figure}

	Assuming that a term like \eqref{RSET} continues to dominate the RSET even after backreaction has become significant, we can extrapolate the movement of the inner horizon further along the evaporation process. For example, for a classical background in which the surface gravity increases exponentially as in the Reissner-Nordström case, the extrapolated self-consistent solution can be observed in fig. \ref{f4}. The inner horizon initially moves inward while the classical background still dominates, but when the semiclassical perturbation has enough time to accumulate it produces an outward bounce and a rapid \textit{inflationary} extinction of the trapped region from the inside.
	
	\section{Conclusions and discussion}\label{secconcl}
	
	The inner horizon and the process of mass inflation are components of black hole physics which are indispensable when constructing a global picture of the spacetimes of these objects, from their formation to their ultimate fate. However, semiclassical analyses of this full picture have thus far been incomplete. Past works have mostly focused on calculating the RSET in background geometries with a Cauchy horizon in order to see how generic its divergent behaviour is on this horizon. This observation is then typically used to suggest that strong cosmic censorship could be saved (from its classical problem of extensions past the weak null singularity) through semiclassical backreaction, as generic initial conditions would seemingly change the Cauchy horizon into a strong singularity.  Less often, it has been indeed appreciated~\cite{BalbinotPoisson93,Ori2019,Zilberman2022} that semiclassical backreaction might lead to ``defocusing" or ``expansion" of the Cauchy horizon to large radii due to the addition of negative mass (essentially coinciding with our result). However, the effect this could have on a dynamically formed trapped region at finite times has not been addressed. Crucially, the idea that the trapped region might have a finite lifetime has not been incorporated into these analyses.
	
	In this work, we have provided a first glimpse into what a complete picture of black hole formation and evolution may look like in semiclassical physics. To this end, we have first used a simple shell-based construction to understand better the classical dynamics of the trapped region in a realistic black hole formation scenario. That has lead to interesting results in its own right: on the one had, the mass inflation instability being triggered depends strongly on the mass to charge (and, by extension, angular momentum) ratio of the infalling matter. In more general black hole constructions with an inner horizon (e.g. singularity-free black holes) the necessary condition for this instability appears to be even more strongly model-dependent, as it comes down to how the infalling perturbations affect the inner horizon position in time.
	
	In the semiclassical picture, for a static (or nearly static) inner horizon not undergoing classical mass inflation, our previous work \cite{Barcelo2021} has shown that backreaction from the RSET has a tendency to induce an inflationary instability, wherein this horizon is displaced in an outward direction in an exponential manner, reminiscent of (a negative version of) the mass inflation effect itself. Extrapolating from this tendency, the trapped region can be assumed to have a very short lifetime, acting rather as a transient between collapse and the formation of a final horizonless or extremal configuration.
	
	Here we have extended this analysis to black holes which, in the absence of other regularising mechanisms, are undergoing classical mass inflation. We first performed a perturbative analysis, which by itself already suggested that semiclassical effects always become dominant at some point in the evolution, as they depend on the surface gravity of the inner horizon which grows exponentially during mass inflation. We then obtained a simple set of self-consistent solutions which, in the absence of the RSET source, can reproduce the structure of a mass inflation geometry, but in a complete semiclassical treatment reveal a very different behaviour. They present a tendency for the semiclassically induced outward push on the inner horizon to accumulate exponentially quickly, leading to an evolution similar to the case of a classically static inner horizon.
	
	\begin{figure}[t]
		\centering
		\includegraphics[scale=1.1]{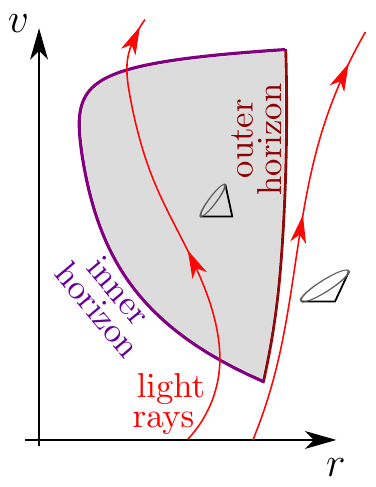}
		\caption{Qualitative picture of the extrapolated evolution of a trapped region in semiclassical physics.}
		\label{f5}
	\end{figure}
	
	Within the framework established here, the semiclassical evolution of black holes could have one of three outcomes. First, we must admit that the standard picture remains a possibility. Classical mass inflation may continue to dominate the dynamics around the inner horizon in later parts of the evolution, where the approximations we have used to estimate semiclassical backreaction cease to be accurate. In this case Hawking evaporation of the outer horizon would dominate the first part of the semiclassical evolution, up to the point at which the mass-inflated region (the upper part of which need not be close to the origin) is revealed to the external universe, where a more detailed analysis would be necessary. The physics of the inner horizon plays a secondary role in this picture until very late times.
	
	However, our present results point to two alternative possibilities which could be realized in a fully self-consistent semiclassical evolution. The first one is that the inner horizon moves outward due to backreaction from the RSET, but only up to the point at which it meets the outer one, leaving an extremal black hole remnant. The second alternative, represented in fig.~\ref{f5}, is that the trapped region disappears completely. The dissipation from this transient process (or several iterations thereof) could bring matter to the initial conditions necessary for the formation of semiclassically self-sustained horizonless black hole mimickers \cite{Visser2008,Barcelo2019,Carballo-Rubio2017,Arrechea2021}. In both these possibilities, the dynamical evolution of the inner horizon plays an essential role. Obviously, further work is necessary to fully probe the viability of these scenarios. However, our analyses already highlight that these not much contemplated possibilities deserve full attention.

	\section*{Acknowledgements}
	Financial support was provided by the Spanish Government through the projects   PID2020-118159GB-C43 and PID-2020-118159GB-C44, and by the Junta de Andalucía through the project FQM219. VB is funded by the Spanish Government fellowship FPU17/04471. CB acknowledges financial support from the State Agency for Research of the Spanish MCIU through the ``Center of Excellence Severo Ochoa'' award to the Instituto de Astrof\'{\i}sica de Andaluc\'{\i}a (SEV-2017-0709). RCR acknowledges financial support through a research grant (29405) from VILLUM fonden.

	\nocite{*}
	\bibliography{Bibliografia}
	\bibliographystyle{ieeetr}

\end{document}